\pdfoutput=1

\documentclass{elsart}

 \usepackage{graphicx}
\usepackage{amssymb}

\begin{document}

\begin{frontmatter}



\title{Evolution of community structure in the world trade web}


\author{Irena Tzekina\corauthref{cor}},
\corauth[cor]{Corresponding author. Address: Department of Mathematics, Dartmouth College, Hanover, NH 03755}
\ead{Irena.Tzekina@Dartmouth.edu}
\author{Karan Danthi},
\ead{Karan.D.Danthi.07@alum.dartmouth.org}
\author{Daniel N. Rockmore}
\ead{rockmore@gauss.dartmouth.edu}
\address{Department of Mathematics, Dartmouth College, Hanover, NH 03755}

\begin{abstract}
In this note we study the bilateral merchandise trade flows between 186 countries over the 1948-2005 period using data  from the International Monetary Fund. We use Pajek to identify network structure and behavior across thresholds and over time. In particular, we focus on the evolution of trade ``islands" in the a world trade network in which countries are linked with directed edges weighted according to fraction of total dollars sent from one country to another. We find mixed evidence for globalization.
\end{abstract}

\begin{keyword}
network, community structure, small world, world trade web
\PACS 89.75.Fb (Structures and organization in complex systems)
\end{keyword}
\end{frontmatter}

\section{Introduction}
\label{sec:intro}
In the past decade or so network analysis has become an important tool in many different disciplines, including finance, economics, sociology, and biology   for the quantitative understanding of relationships (see \cite{newman-survey} for a thorough survey of applications and many references). In this paper we use the tools of network analysis to investigate international trade relationships. More specifically, using IMF data \cite{data2} we look at the bilateral merchandise trade flows between 178 nations over the years 1948-2005. We consider the network of nations in which an edge goes from nation $i$ to nation $j$ if and only if the fraction of total trade of $i$ due to $j$ exceeds a given threshold.

In the language of \cite{Fagiolo}, our primary results fall into the category of  a ``descriptive" use of  network theory in the investigation of international trade - that is, we use network-based analyses to reveal and illuminate certain structures or patterns in international  trade. In particular, we consider the formation of  ``islands" of our directed network (obtained as described above) and see over time, an evolution away from two large trading islands centered on the postwar superpowers of the UK and US to a fairly heterogeneous ``archipelago" of trade, seeming to reflect some aspect of globalization, at least from the point of view of finding multiple centers of trade. Some islands seem highly regionalized, other show vestiges of old colonialism, while others seem a little more difficult to characterize. Thus do we identify an evolution of the community structure in the WTW from a few major trading hubs to a more decentralized topology for international trade. 
Geopolitical events (e.g., trade treaties and well-known political alliances such as the creation of a trade block corresponding to members of the former Soviet Union) are also naturally exposed. 

Our work falls within the growing body of research now devoted to analyzing the so-called  ``world trade web" (WTW) (see e.g., \cite{Gar2,Li,Serrano1}. Generally, this research  examines the properties of  a network in which the nodes are countries and edges go between nodes based on some metric related to trade. This sort of mathematical context has enabled researchers to draw interesting conclusions regarding the world economy. Of particular relevance to this paper is the  paper \cite{kastelle} in which network theory is used to evaluate aspects of the hypothesis of ``globalisation" or the existence of a global economy. Also of relevance is the paper  \cite{Garl} relating the evolution of the topology of the WTW and the distribution of GDP. Our interest is rather in the evolution of the community structure of the WTW as well as in the set of network hubs.  (Recall that a community in a network is identified as a collection of nodes that have a dense set of edges among each other, but a relatively few number of edges to other 
densely connected sets.) Our observation of decentralization is in agreement with a different longitudinal network analysis found in 
\cite{kim}.

\section{Data and Methodology - Islands in our World Trade Web}

Our data on bilateral trade comes from the ÒDirection of TradeÓ CD developed by the International Monetary Fund \cite{data2}. The dataset includes bilateral merchandise trade data for 186 countries over the 1948-2005 period. The data is recorded in real dollar terms and hence is adjusted for inflation. However, using this data directly would bias our network towards trade linkages involving affluent countries as they trade with other countries in larger quantities. To eliminate this bias, we compute the value of each bilateral trade flow as a fraction of a countryÕs total trade volume. This gives a directed network based on countries weighted by fraction of total trade from source to target.

In order to analyze the structure and behavior of the network through time, we use the notion of  ``islands," which is implemented  in Pajek \cite{pajek}, as a means to identify communities and hubs.  Islands correspond to connected components in which the weights of the arcs within the community are significantly larger than those outside the community. We generate these islands based on a $5\%$ threshold\footnote{I.e., we remove from the full directed network all trade links that represent less than $.05$ of total trade from a given country.} , so that we can capture communities of the most significant trading links, without compromising on the accuracy of our network. Thus, in order to be connected to an island, the country must dedicate significantly more of its trade volume to other members of the island than to nations outside of it; unless this differential exists, the country will not appear in the island. The resulting figure signifies that the recipient of an arc is a major trading partner of the source country at the $5\%$
   level and that trade with the source country significantly outweighs trade with other countries in the network. 
   These $5\%$
    threshold island figures for 1950, 1960,1970,1980,1990, 1999 and 2005 are depicted in Figures 1-3. 

These graphs enable us to identify  communities and hubs, which we define as a country that is a significant trading partner for many, if not most, members of the cluster. Hubs are nations with  large in-degree. 

\section{The Evolution of Island Structure and an Interpretation}

This island-based approach seems to provide a network theoretic language for illustrating various aspects of  post World War II international political and economic history.  
A thumbnail historical summary might go something like this: In 1950, World War II had recently come to an end, and many nations were recovering from the costs associated with the conflict.  
We see two large islands, centered upon two of the post-war superpowers,  the US and the UK,  that mainly show  regional or (historically) colonial economic influence. Also note the relatively small number of nodes - few countries have significant trade with any particular partner.

The escalation of the Cold War in the 1960s coincides with the emergence of  an East-European cluster in our islands graph for 1960. This cluster consists of Bulgaria, Poland and Hungary. In contrast, a Western multi-hub community, comprised of Italy, Germany and France, themselves major trading partners for other nations, also forms. By 1970 Germany becomes a significant trading partner for other Western European hubs such as France and Italy. This development could be tied to the improved political and economic ties of West Germany, as well as its strong economic performance. However, the Berlin Wall falls in 1989, and Germany is once again unified. As a result, in the 1990 network some former Communist nations, which probably still maintain economic ties with former East Germany, are now in the cluster. By the late 1990s, the wane of Communism and the fall of the Soviet Union 
gives rise to a ``transition-economy" cluster  of former Soviet states that forms around Russia.  By 2005, Russia has recovered from the economic problems she experienced in the late 1990s, and her significance in the network grows. Russia is now one of the largest hubs in the network and serves as the major trading partner for many East European nations. 
The formation and expansion of the European Union  may have also contributed to the expansion of the Western European cluster. 

The move of Africa away from colonialism toward independence marks another important political transition that occurs during the 1960s. This phenomenon seems manifested in the 1970 network in two ways. Firstly, African clusters such as that of Zambia, Namibia and Zimbabwe, separate from the colonializing country (i.e the hub), emerge. Secondly, some African nations simply move away from the parent country. One example of this phenomenon is South AfricaÕs separation from the United Kingdom and its disappearance form the network, illustrating that the country did not significantly trade with anyone else in the islands graph above the $5\%$
 threshold. Another is the Democratic Republic of Congo's move from being one degree of separation from France to being two degrees of separation away. By 1980, South Africa rejoins the network to become a part of a mini cluster, as have other African nations, such as Ethiopia and Tanzania. Benin also moves away from France and makes an economic tie with India. Finally, following the end of Apartheid in 1994, South Africa becomes a hub for African trade in the 1999 network; this trend continues through 2005.

   \section{Concluding Remarks}

The evidence for or against globalization is mixed, at least from the point of view of the Islands decomposition.  On the one hand,  we see many more countries that trade significantly (i.e.,  at more than $\% 5$) in 2005 than in 1950, indicating that trade was more spread out in 1950 than in 2005. On the other hand, in 2005 there are many more islands, an indication that more trade centers exist - hence from the point of view of significant markets, more have evolved. In summary, we seem to be seeing a consolidation of insignificant trading, followed by diversification. This network-based mixed view on globalization echoes previous work \cite{kastelle}. In the WTW ``archipelago" of 2005 we see a heterogeneity in the islands, although by and large, regional influences seem to play a role and we do generally witness a diminishing of colonial influence.  This can both be a reflection of local trade agreements as well as (presumably) cheaper transportation costs.

   As a topic of further analysis, it would interesting to look at how the network of trade in services compares with our network, and thus expand on the present analysis. 
   Because the scope of our analysis is so broad, we can only make a few conclusions about the evolution of trading patterns in specific regions. It would be worthwhile to carry out region-specific analyses (e.g. the European Union).



\

\begin{figure}
\centering
\includegraphics[height=70mm]{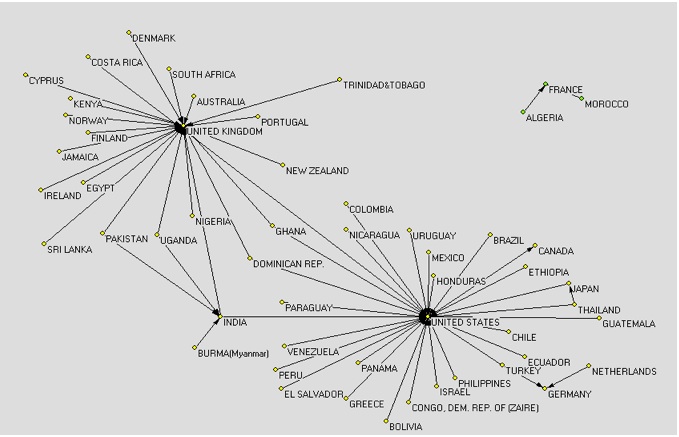}
\includegraphics[height=70mm]{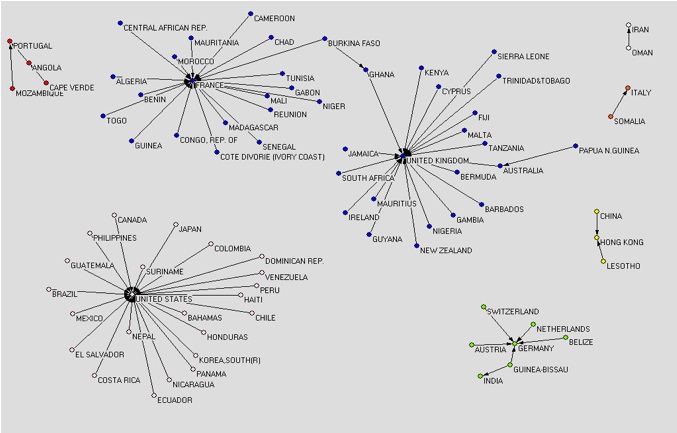}
\includegraphics[height=70mm]{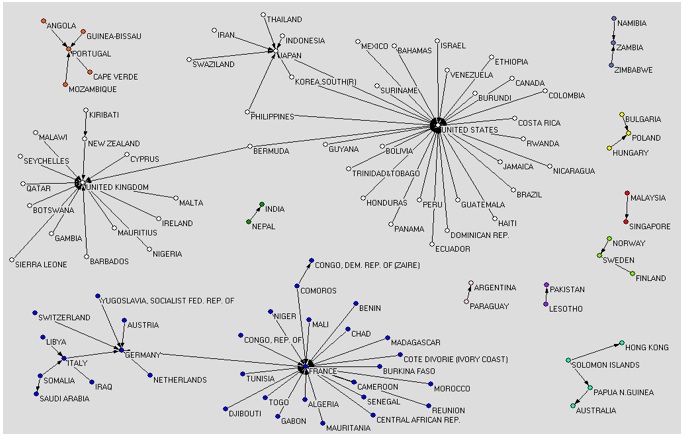}
\caption{Top to bottom: Island structure   for  1950, 1960, and 1970 WTW. }
\end{figure}

\begin{figure}
\centering
\includegraphics[height=70mm]{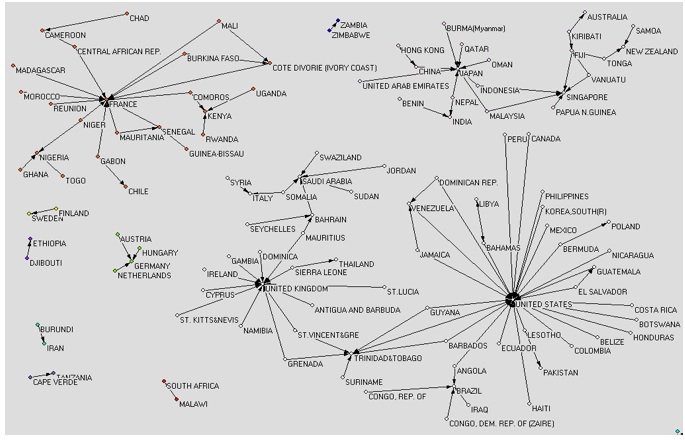}
\includegraphics[height=70mm]{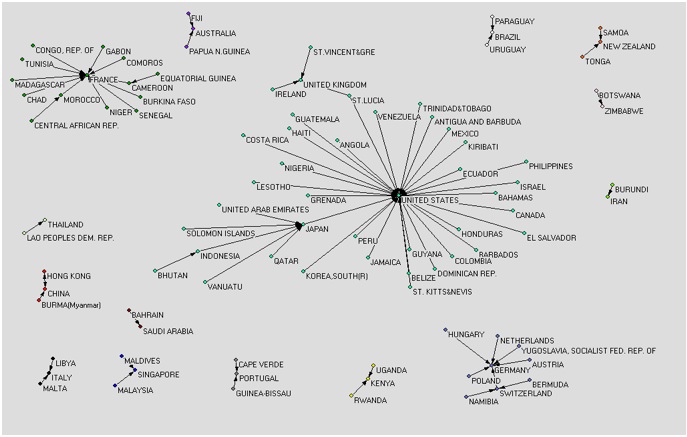}
\includegraphics[height=70mm]{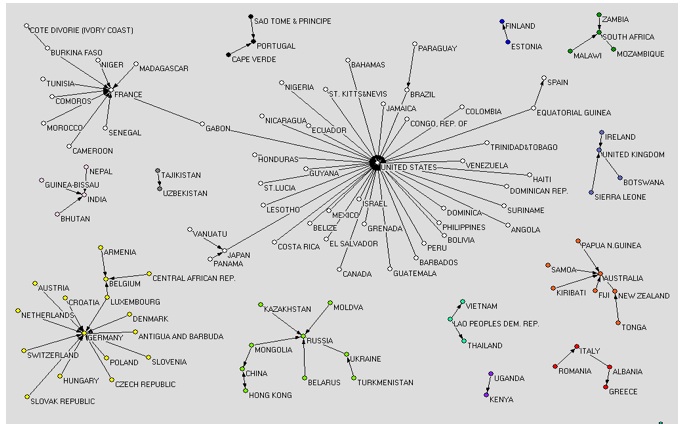}
\caption{Top to bottom: Island structure   for  1980, 1990, and 1999 WTW.  }
\end{figure}

\begin{figure}
\centering
\includegraphics[height=80mm]{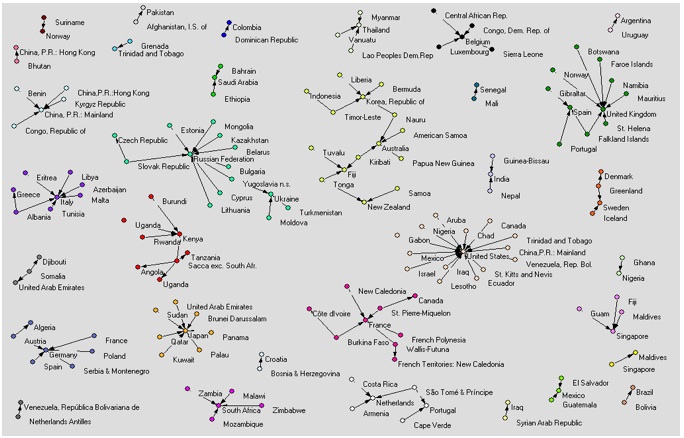}
\caption{Islands structure  for the 2005 WTW. }
\end{figure}


\begin{thebibliography}{00}






\bibitem{data2}
ESDS International,IMF Direction of Trade Statistics. CD-ROM.Washington, D.C.:IMF ,March 2007.





\bibitem{Fagiolo}
G. Fagiolo, J. Reyes, S. Schiavo, The evolution of the world trade web, 
Tech. Rep. 2007-16, Sant\' Anna School of Advanced Studies, LEM Working Paper, 2007.

\bibitem{Garl}
D. Garlaschelli, T. Di Matteo,  T.  Aste,  G. Caldarelli, M. I. Loffredo,  ``Interplay between topology and dynamics in the World Trade Web," Proceedings of the 5th conference on Applications of Physics in Financial Analysis (APFA5), 29 June - 1 July 2006, Torino (ITALY), arXiv:physics/0701030v1, 2007 

\bibitem{Gar2}
D. Garlaschelli, M. Loffredo, Fitness-dependent topological properties of the world trade web, Physical Review Letters {\bf 93} (2004) 188701.


\bibitem{kastelle}
T. Kastelle, Tim, J. T. Steen, P. W. Liesch, 
Measuring globalisation: An evolutionary economic approach to tracking the evolution of international trade,
in Proceedings of DRUID Summer Conference,  June,  2006.

\bibitem{kim}
S. Kim, E. H. Shin, 
A Longitudinal Analysis of Globalization and Regionalization in International Trade: A Social Network Approach
Social Forces - Volume 81, Number 2, December 2002, pp. 445-471

 \bibitem{Li}
 X. Li, Y. Y. Jin, G. Chen, 
 Complexity and synchronization of the world trade web,
 Physica A: Statistical Mechanics and its Applications 
Volume 328, Issues 1-2, 1 October 2003, Pages 287-296.

\bibitem{newman-survey}
M. E. J. Newman, Structure and function of complex networks, {\textit  SIAM Review}, 45(2):167-256, 2003.

\bibitem{pajek}
http://vlado.fmf.uni-lj.si/pub/networks/pajek/

\bibitem{Serrano1}M. A. Serrano, M. Bogu\~{n}\'{a}, A. Vespignani,
Patterns of dominant flow in the world trade web, ArXiv e-prints,  arXiv:0704.1225 (2007). 


\end{thebibliography}
\end{document}